\documentstyle[aps,prl,epsfig,twocolumn]{revtex}
\newcommand{\beeq}{\begin{equation}}
\newcommand{\eneq}{\end{equation}}
\newcommand{\beeqar}{\begin{eqnarray}}
\newcommand{\eneqar}{\end{eqnarray}}
\begin{document}
\twocolumn[\hsize\textwidth\columnwidth\hsize\csname
  @twocolumnfalse\endcsname
\title{
Information cloning of harmonic oscillator coherent 
states and its fidelity.
}
\author{N.D. Hari Dass \dag } 
\address{The Institute of Mathematical Sciences, Chennai - 600 113, INDIA}
\author{Pradeep Ganesh \ddag} 
\address{Indian Institute of Technology, Chennai, INDIA}
\maketitle
\begin{abstract}
We show that in the case of  unknown {\em harmonic 
oscillator coherent states} it is possible to
achieve what we call {\it perfect information cloning}.
By this we mean that it is still possible
to make arbitrary number of copies of a state
which has {\it exactly} the same
information content as the original unknown coherent
state. By making use of this {\it perfect information
cloning} it would be possible to estimate the original
state through measurements and make arbitrary number
of copies of the estimator. We define the notion of a
{\em Measurement Fidelity}. We show that this information
cloning gives rise, in the case of $1\rightarrow N$, to a {\em distribution} of 
{\em measurement fidelities} whose average value is ${1\over 2}$
irrespective of the number of copies originally made. Generalisations of this to the
$M\rightarrow MN$ case as well as the measurement fidelities
for Gaussian cloners are also given.
\end{abstract}
\vskip 2pc] 
\section{Introduction}
{\label{intro}}
The no-cloning theorem 
expresses the inability to copy an
{\em unknown} quantum state. The original version 
\cite{noclone} 
of the
theorem invoked the
principle of linear superposition. 
The
stronger restriction on cloning in fact comes from the {\em
unitarity} of the cloning transformation itself. 
Accordingly, only mutually orthogonal states can be cloned universally.
For {\em class-dependent} cloning the corresponding statement is
that states belonging to the same class should still be mutually
orthogonal but states belonging to distinct classes need not be so.

This precludes perfect cloning of even subclasses of  harmonic 
oscillator coherent states. 
Cerf et al \cite{optimal1} have shown that there is an 
optimal fidelity to cloning coherent states.
Many significant results regarding
optimality of cloning of coherent states have appeared recently in the
literature \cite{optimal1,optimal2,optimal3}. Proposals for optical
implementations have also been made \cite{optical}.

In this paper we wish to present an alternate route to the
question of cloning coherent states. We show that
it is possible to make arbitrary number of
copies of coherent states with exactly the same information content as the
original unknown state. Complete
information about a coherent state is contained in the complex coherency
parameter $\alpha$. Thus by information cloning what we mean is the
ability to make arbitrary number of copies of coherent states 
whose coherency parameter is $c(N)\alpha$ where $\alpha$ is the coherency
parameter of the {\em unknown} coherent state and $c(N)$ is a {\em known} constant
depending on the number of copies made.

We consider $1+N$ systems of harmonic oscillators whose
creation and annihilation operators are the set $(a,a^\dag),
(b_k,b_k^\dag)$ (where the index $k$ takes on values $1,..,N$)
satisfying the commutation relations
\beeq \label{4}
[a,a^\dag]~=~1;~~[b_j,b_k^\dag]~=~\delta_{jk};~~[a,b_k]~=~0;~~[a^\dag,b_k]~=~0
\eneq
Coherent states parametrised by a complex number are given by
\beeq \label{5}
|\alpha >~=~D(\alpha)~|0 >
\eneq
where $|0>$ is the ground state and the unitary operator $D(\alpha)$ is given by
\beeq \label{6}
D(\alpha)~=~e^{\alpha~a^\dag~-\alpha^*~a}
\eneq
Let us consider a {\it disentangled} set of coherent states
$|\alpha >|\beta_1 >_1|\beta_2 >_2...|\beta_N >_N$ and consider the action of the unitary transformation 
\beeq \label{7}
U = e~~^ {~t(a^\dag\sum_j \kappa_j b_j - a\sum_j \kappa_j^* b_j^\dag)}
\eneq
By an application of the Baker-Campbell-Hausdorff 
identity and the fact that $U|0>|0>_1..|0>_N=|0>|0>_1..|0>_N$ it is easy
to see that the resulting state is also a disentangled set of coherent
states expressed by
\beeq \label{8}
|\alpha^\prime>|\beta_1^\prime>_1..|\beta_N^\prime>_N~~
=~~U~~ |\alpha>|\beta_1>_1..|\beta_N>_N~~
\eneq
It is useful to transform the problem to the case where all $\kappa_j$ are real through a redefinition of the creation and
annihilation operators. Denoting $\kappa_j=r_je^{i\delta_j}$
we can introduce the tranformation
\beeq \label{9}
\tilde a~=~a~~~~\tilde b_j ~=~ e^{i\delta_j} b_j~~~~~ \tilde b_j^\dag ~=~ e^{-i\delta_j} b_j^\dag
\eneq
This leaves the commutation relations of eqn (1) unchanged
\beeq \label{10}
[\tilde a,\tilde a^\dag]~=~1;~~[\tilde b_j,\tilde b_k^\dag]~=~\delta_{jk};~~[\tilde a,\tilde b_k]~=~0;~~[\tilde a^\dag,\tilde b_k]~=~0
\eneq
The parametrisation of the coherent states are correspondingly redefined as
\beeq \label{11}
\tilde\alpha~=~\alpha~~~~~~\tilde\beta_j~=~e^{-i\delta_j}~\beta_j
\eneq
The unitary operator $U$ takes the form
\beeq \label{12}
\tilde U = e~~^ {~t(\tilde a^\dag\sum_j r_j \tilde b_j - \tilde a\sum_j r_j \tilde b_j^\dag)}
\eneq
The initial state is
\beeq \label{13}
|I>~=~D(\tilde \alpha)~D(\tilde\beta_1)_1...D(\tilde\beta_N)_N~~
|0>|0>_1..|0>_N
\eneq
Defining
\beeq \label{14}
\tilde a(t)~=~\tilde U~\tilde a~\tilde U^\dag~~~~~~~\tilde b_j(t)~=~\tilde U~\tilde b_j~\tilde U^\dag
\eneq
one easily gets the differential equations
\beeq \label{16}
{d\over dt}~\tilde a(t)~=~-\sum_j~r_j\tilde b_j(t)~~~~
{d\over dt}~\tilde b_j(t)~=~r_j\tilde a(t)
\eneq
The solutions to these eqns are straightforward to find:
\beeqar \label{17}
\tilde a(t)~&=&~{\cos rt}~\tilde a~-~\sum_j~{r_j\over r}{\sin rt}~\tilde b_j\nonumber\\
\tilde b_j(t)~&=&~{r_j\over r}{\sin rt}~ \tilde a+\sum_k~
\tilde M_{jk}(t)~\tilde b_k
\eneqar
where $r=\sqrt(\sum_j~r_j^2)$ and 
\beeq \label{18}
\tilde M_{jk}~=~\delta_{jk}-{r_jr_k\over r^2}(1-{\cos rt})
\eneq
This transformation induces a transformation on the parameters
$(\tilde\alpha,\tilde\beta_j)$ which can be represented by the
matrix $\tilde {\cal U}$ i.e ${\tilde \alpha}_a(t)=
{\tilde{\cal U}}_{ab}{\tilde \alpha}_b$. We have introduced 
the notation 
$\tilde\alpha_a$ with $a=1,...,N+1$
such that $\tilde\alpha_1 = \tilde\alpha,\tilde\alpha_k=\tilde\beta_{k-1}(k\geq 2)$ 
and a similar
notation for the $(\alpha,\beta_j)$ with ${\cal U}$
as the corresponding matrix. Then we have
\beeq \label{19}
\tilde{\cal U}_{1a} = \left(\begin{array}{ccccc}
                             {\cos rt}&{r_1\over r} {\sin rt}
			     & ..&..&{r_N\over r} {\sin rt}
			     \end{array}\right)
\eneq
\beeq \label{20}
\tilde{\cal U}_{ab} = -{r_{a-1}\over
r}~{\sin rt}~\delta_{b1}+(1-\delta_{b1})\tilde M_{a-1,b-1}
\eneq
where eqn (\ref{20}) is defined for $a\geq 2$. 
Equivalently
 \begin{equation} \label{21}
         \tilde{\cal U} = \left( \begin{array}{ccccc}
 {\cos rt}&{r_1\over r}~{\sin rt}  &..  &..  &{r_N\over r}~{\sin rt}\\
 -{r_1\over r}~{\sin rt}&\tilde M_{11}&.. &.. &\tilde M_{1N}\\
 .. &.. &.. &.. &\\
 .. &.. &.. &.. &\\
 -{r_N\over r}~{\sin rt}&\tilde M_{N1}&.. &.. &\tilde M_{NN}\\

	     \end{array} \right )
 \end{equation}
 It is easy to show that the matrix $\tilde {\cal U}$ is {\it Orthogonal}
 satisfying 
 \beeq \label{22}
 \sum_a~\tilde {\cal U}_{ab}~~\tilde{\cal U}_{ac}~=~\delta_{bc}
 \eneq
 The orthogonality of this matrix can be demonstrated directly on
 physical grounds. It can easily be seen that $ U$(respectively 
$ \tilde U$) commutes with $a^\dag a+\sum_k b_k^\dag b_k$
(respectively 
$\tilde a^\dag \tilde a+\sum_k \tilde b_k^\dag \tilde b_k$). This
implies that
\beeqar \label{23}
|\tilde\alpha(t)|^2+\sum_k |\tilde\beta_k(t)|^2 &=& |\tilde\alpha|^2+\sum_k |\tilde\beta_k|^2 \nonumber\\
|\alpha(t)|^2+\sum_k |\beta_k(t)|^2 &=& |\alpha|^2+\sum_k |\beta_k|^2 
\eneqar
This means that $\tilde{\cal U},{\cal U}$ are {\it Unitary matrices}.
Since
$\tilde{\cal U}$ is {\it real} it is {\it Orthogonal}. The orthogonality
of $\tilde{\cal U}$ leads to another very important quadratic invariant
in addition to the ones stated in eqn(19), namely,
\beeq \label{24}
\tilde\alpha^(t)^2+\sum_k \tilde\beta_k(t)^2 = \tilde\alpha^2
                                                +\sum_k \tilde\beta_k^2 
\eneq
In fact for two independent sets of coherent state parameters
$(\alpha,\vec\beta),(\eta,\vec\xi)$ we have the invariants
\beeqar \label{25}
\tilde\alpha(t)~\tilde\eta(t) +\sum_k \tilde \beta_k(t)~\tilde\xi(t)_k
~&=&~
\tilde\alpha~\tilde\eta +\sum_k \tilde \beta_k~\tilde\xi_k\nonumber\\
\tilde\alpha^*(t)~\tilde\eta(t)~~+\sum_k \tilde \beta_k^*(t)~\tilde\xi_k(t)
~&=&~
\tilde\alpha^*~\tilde\eta~~+\sum_k \tilde \beta_k^*~\tilde\xi_k
\eneqar
Inverting the transformations in eqn (\ref{11}) it is \ref{11} straightforward to
obtain
 \begin{equation} \label{21.2}
         {\cal U} = \left( \begin{array}{ccccc}
 {\cos rt}&{r_1\over r}~{e^{-i\delta_1}\sin rt}  &..  &..  &{r_N\over r}~{e^{-i\delta_N}\sin rt}\\
 -{r_1\over r}~{e^{i\delta_1}\sin rt}&M_{11}&.. &.. &M_{1N}\\
 .. &.. &.. &.. &\\
 .. &.. &.. &.. &\\
 -{r_N\over r}~{e^{i\delta_N}\sin rt}&M_{N1}&.. &.. &M_{NN}\\

	     \end{array} \right )
 \end{equation}
 where
 \beeq \label{26}
 M_{jk}~=~\delta_{jk}-e^{i\delta_j-i\delta_k}{r_jr_k\over r^2}(1-{\cos
 rt})
 \eneq
 The corresponding quadratic invariants are
\beeqar \label{27}
\alpha(t)~\eta(t) +\sum_k e^{-2i\delta_k} \beta_k(t)~\xi(t)
~&=&~
\alpha~\eta +\sum_k  e^{-2i\delta_k}\beta_k~\xi\nonumber\\
\alpha^*(t)~\eta(t) +\sum_k  \beta_k^*(t)~\xi(t)
~&=&~
\alpha^*~\eta +\sum_k  \beta_k^*~\xi
\eneqar
With these results we first illustrate the hurdles to be overcome in
cloning the harmonic oscillator coherent states. For this purpose let us
look at the case $N=1$. In this case the state obtained after applying the
unitary transformation $U$ is
\beeqar \label{28}
\alpha(t)~&=&~{\cos rt}~ \alpha + e^{-i\delta}~ {\sin rt}~ \beta\nonumber\\
\beta(t)~&=&~{\cos rt}~ \beta - e^{i\delta}~ {\sin rt}~ \alpha
\eneqar
Here the best that can be achieved is to set ${\sin rt}=1$ and remove
known phases through the unitary operator $e^{i\gamma~a^\dag a}$ and one
sees that it only amounts to {\it swapping} but not cloning. Next, in the
general $N$ case, the coefficients of $\alpha$ in all $\beta_k(t)$ must be
made to have the same magnitude implying $r_1=r_2=.....=r_n$. With the
choice $\beta_1=\beta_2=..=\beta_N$ one gets
\beeq \label{29}
\beta_k(t)=-e^{i\delta_k}{{\sin rt}\over {\sqrt N}}\alpha
\eneq
With the optimal choice of ${\sin rt}=-1$ and using 
appropriate unitary
transformations to remove known phases one gets N copies of the state
$|{\alpha\over {\sqrt N}}>$. 
\subsection{Information cloning}
Thus we are able to produce
N-copies {\em not} of the original state $|\alpha>$ but of a state of the
form $|{\alpha\over\sqrt N}>$ which has the {\bf same information content}
as $|\alpha>$ in the sense that a complete determination of the latter
is equivalent to a complete determination of the former. This is what we
would like to call {\it cloning of information} in contrast to 
{\it cloning of the quantum state itself}. It is quite plausible that in many 
circumstances of interest cloning in this more restricted sense may
suffice.

Superficially this may appear to be a triviality in the sense that one
can always apply known unitary transformations on unknown quantum states
to produce states with the same information content in the sense used 
above. But what is {\it nontrivial} in our construction is that 
{\em arbitrary} number of copies of such information-equivalent states 
can be produced.

Fidelity of cloning was interpreted by \cite{optimal1} as
$<\alpha|\rho_1|\alpha>$ where $|\alpha>$ is the
unknown coherent that was cloned and $\rho_1$ is the
one-particle reduced density matrix of the output. In
the Gaussian-cloners of the type considered in 
\cite {optimal1} there are N-copies of $\rho_1$ which
are all {\em mixed} states. In contrast our {\em
information cloning} produces N-copies which are
{\em pure} states. We call the fidelity introduced by
\cite{optimal1} {\it Overlap Fidelity}. This overlap
fidelity for our information cloning is
\beeq
F_{info}^{overlap} = e^{-|\alpha|^2(1-{1\over\sqrt N})^2}
\eneq
Not only can this be very small it is also not {\em universal}.

We introduce another notion of fidelity which we call
{\em Measurement Fidelity} by which we mean the best
reconstruction of the original unknown state that can
be achieved through actual measurements performed in
some optimal way. Before proceeding we give the
formula for the optimal overlap fidelity of 
$N\rightarrow M$
cloning of coherent states \cite{optimal1}
\beeq \label{overlap}
F_{N,M}^{overlap} = {MN\over MN+N-M}
\eneq
We now propose using the
copies of the information-equivalent states to estimate the parameter
$\alpha$. Normally when the available number of copies of a state is very large, one
one can estimate the state quite accurately and use that to create arbitrary
number of clones of the original coherent state. However, in our proposal
for information cloning even though the number of copies $N$ can be
arbitrarily {\em large}, the coherency parameter given by 
${\alpha\over\sqrt N}$ becomes {\em arbitrarily small} while the
{\em variances} in $\alpha$ remain the same as in the original state.
This raises the question as to how best the original state can be reconstructed
and about the statistical significance of our information cloning procedure.

On introducing momentum and position operators $\hat p,\hat x$ through
\beeq \label{43}
{\hat x} = {(a+a^{\dag})\over\sqrt 2}~~~~ {\hat p} = {(a-a^{\dag})\over\sqrt 2i}
\eneq
the {\em probability distributions } in position and momentum representations 
are given
by
\beeqar \label{dist}
|\psi_{clone}(x)|^2& = &{1\over\sqrt\pi} e^{-(x-\sqrt {2\over N} \alpha_R)^2}\nonumber\\ 
|\psi_{clone}(p)|^2& = &{1\over\sqrt\pi} e^{-(p-\sqrt {2\over N} \alpha_I)^2} 
\eneqar
Let us distribute our $N$-copies into two groups of 
$N/2$ each and use one
to estimate $\alpha_R$ through position measurements 
and the other to
estimate $\alpha_I$ through momentum measurements.
Let $y_N$ denote the average value of the position
obtained in $N/2$ measurements and let $z_N$ denote the
average value of momentum also obtained in $N/2$
measurements. The {\em central limit theorem} states
that the probability distributions for $y_N,z_N$ are
given by
\beeqar \label{45}
f_x(y_N)&=&\sqrt{N\over 2\pi}e^{-{N\over 2}(y_N-\sqrt{2\over N}\alpha_R)^2}\nonumber\\
f_p(z_N)&=&\sqrt{N\over 2\pi}e^{-{N\over 2}(z_N-\sqrt{2\over N}\alpha_I)^2}
\eneqar
The {\em estimated} value of $\alpha$ is
\beeq \label{46}
\alpha_{est} = {y_N+iz_N\over\sqrt 2}\sqrt N
\eneq
The {\em measurement fidelity} $F^{meas}$ can be 
understood as the quantity
$|<\alpha|\alpha_{est}>|^2$:
\beeq \label{47}
F^{meas} = e^{-|\alpha - \alpha_{est}|^2}
\eneq
The probability distribution for $F$ is given by
\beeq \label{48}
p(F)dF=\int dz_Ndy_N~\delta(z_N^2+y_N^2-{2\over N}|\alpha_{est}|^2)f_x(y_N)f_p(z_N)
\eneq
It is straightforward to show that
\beeq \label{49}
p(F)dF=dF
\eneq
Consequently the average value of $F^{meas}$ is 
$$
\bar F_{1,N}^{meas}=1/2
$$
Now we generalise our results to the $M'\rightarrow N'$ case. We start with $M$ copies and let each copy be 
information cloned to $N$ copies so we have 
$MN$ copies finally.

The position and momentum distributions are still given
by eqn (\ref{dist}) but now $NM/2$ measurements are
carried out for position and momentum. Consequently
\beeqar \label{central}
f_x(y_{MN})&=&\sqrt{MN\over 2\pi}e^{-{MN\over 2}(y_{MN}-\sqrt{2\over {N}}\alpha_R)^2}\nonumber\\
f_p(z_{MN})&=&\sqrt{MN\over 2\pi}e^{-{MN\over 2}(z_{MN}-\sqrt{2\over {N}}\alpha_I)^2}
\eneqar
The {\em estimated} value of $\alpha$ is still given by eqn(\ref{46}).
One finally obtains
\beeq
p_{M,MN}(F)dF=  M F^{M-1} dF
\eneq
The average measurement fidelity in this case is 
given by
\beeq \label{mfidel}
{\bar F}^{meas}_{M,MN} = {M\over M+1}
\eneq
This approaches $1$ as $M\rightarrow\infty$.

These fidelities should not be directly compared 
with eqn(\ref{overlap}).
As emphasised by Massar and Popescu \cite{massar} there can
be many notions of fidelities and two schemes
should be compared only with the {\it same} criterion
for fidelity. So we compute the
measurement fidelity for Gaussian cloners.
 Each copy is the Gaussian mixture
\beeq
\rho = \int d^2\alpha {A_{M,MN}\over \pi} e^{-A_{M,MN}|\alpha|^2}
|\alpha_0+\alpha><\alpha_0+\alpha|
\eneq
Where $A_{M,MN}$ given by
\beeq
A_{M,MN} = {MN\over N-1}
\eneq
reproduces eqn (\ref{overlap}).
The position and momentum distributions in the Gaussian mixture are
given by
\beeqar 
p_{Gauss}(x)& =& {1\over\sqrt\pi}{\sqrt {A_{M,MN}\over A_{M,MN}+2}} e^{-(x-\sqrt {2} \alpha_R)^2} \nonumber\\
p_{Gauss}(p)& =& {1\over\sqrt\pi} {\sqrt {A_{M,MN}\over A_{M,MN}+2}}e^{-(p-\sqrt {2} \alpha_I)^2} \nonumber
\eneqar
The analogues of eqns (\ref{central}) are given by
\beeqar 
f_x(y_{MN})&=&\sqrt{MNA_{M,MN}\over 2\pi (A_{M,MN}+2)}e^{-{MN\over 2}(y_{MN}-\sqrt{2}\alpha_{0R})^2}\nonumber\\
f_p(z_{MN})&=&\sqrt{MNA_{M,MN}\over 2\pi (A_{M,MN}+2)}e^{-{MN\over 2}(z_{MN}-\sqrt{2}\alpha_{0I})^2}\nonumber
\eneqar
The estimate for the coherency parameter is
\beeq
\alpha_{est} = {y_{MN}+iz_{MN}\over\sqrt 2}
\eneq
The resulting measurement fidelity distribution is
\beeq
p^{Gauss}_{M,MN}(F)dF=F^{{MNA_{M,MN}\over 2(A_{M,MN}+2)}-1}dF
\eneq
while the average measurement fidelity is
\beeq
{\bar F}^{Gauss}_{M,MN} = {M^2N^2\over M^2N^2+2MN+4N-4}
\eneq

For $M=1,N=2$ the measurement fidelities for Gaussian and Information cloning
are $1/3$ and $1/2$ respectively. For $M=1,N=4$ these become $4/9$ and $1/2$.
For $M=2,N=2$ these are $4/7$ and $2/3$ while for $M=2,N=4$ they become
$16/23$ and $2/3$ respectively.
\subsection{Conclusion}
In this paper we have demonstrated the concept of {\em information
cloning} for harmonic oscillator coherent states. The principal difference
with the {\em Gaussian cloning} of \cite{optimal1,optimal2} is that
in our case the outputs are {\em pure} and {\em disentangled} states. The
coherency parameter for the output states is reduced by the factor ${1\over\sqrt N}$
where $N$ is the number of copies. The variances are unchanged. We have
also introduced the notion of {\em measurement fidelity} which is 
different from the notion of fidelity introduced in \cite{optimal1,optimal2}.
For purposes of comparison we have calculated the measurement fidelities
for Gaussian cloners also. 
In the case of $d$-level
quantum states a formula 
is available giving the fidelity that can be 
achieved given $N$ copies
\cite{fidel1}.
Our formula eqn (\ref{mfidel}) is such a relation for coherent states.


\end{document}